\documentclass[twocolumn,showpacs,amsmath,prb,amssymb]{revtex4}

\usepackage{graphicx}
\usepackage{dcolumn}
\usepackage{bm}

\begin{document}

\title{Two-loop approximation in the Coulomb blockade problem}

\author{I.S.~Beloborodov} \affiliation{Bell Laboratories, Lucent
  Technologies, Murray Hill, New Jersey 07974} \affiliation{Department of
  Physics, University of Colorado, CB 390, Boulder, Colorado 80390}

\author{A.V.~Andreev} \affiliation{Bell Laboratories, Lucent Technologies,
  Murray Hill, New Jersey 07974} \affiliation{Department of Physics,
  University of Colorado, CB 390, Boulder, Colorado 80390}

\author{A.I.~Larkin} \affiliation{Theoretical Physics Institute,
  University of Minnesota, Minneapolis, Minnesota 55455}
\affiliation{L.~D.~Landau Institute for Theoretical Physics, Moscow 117334,
  Russia}

\date{\today}

\begin{abstract}
  We study Coulomb blockade (CB) oscillations in the thermodynamics of a
  metallic grain which is connected to a lead by a tunneling contact with
  a large conductance $g_0$ in a wide temperature range,
  $E_Cg_0^4 e^{-g_0/2}<T<E_C$, where $E_C$ is the charging energy.
  Using the instanton analysis and the renormalization group we obtain the
  temperature dependence of the amplitude of CB oscillations which differs
  from the previously obtained results. Assuming that at $T <
  E_Cg_0^4 e^{-g_0/2}$ the oscillation amplitude weakly depends on
  temperature we estimate the magnitude of CB oscillations in the ground
  state energy as $E_Cg_0^4 e^{-g_0/2}$.
\end{abstract}


\maketitle

\section{\label{Introduction}Introduction and main results}
The study of electron-electron interactions in mesoscopic systems has been
at the focus of experimental and theoretical interest over the past two
decades.  One of the most striking consequences of electron interactions
at low temperatures is the phenomenon of Coulomb
blockade.\cite{Kouwenhoven} For example, thermodynamic quantities of a
metallic grain which is connected by a tunneling contact to a metallic
lead and is capacitively coupled to a gate exhibit oscillatory dependence
on the gate voltage.  This can be observed experimentally by measuring the
differential capacitance of the grain.\cite{Zhitenev}

At low temperatures the Coulomb interaction of electrons in the grain can
be described within the framework of the constant interaction model
\begin{equation}
  \label{interaction} 
  \hat{H}_C=E_C \left(\hat{N}-q\right)^2, 
\end{equation}
where $E_C$ is the capacitive charging energy of the grain, $\hat{N}$ is
the operator of the number of electrons in it, and $q$ is the
dimensionless gate voltage.

For a grain with a vanishing mean level spacing and for a tunneling
contact with a large dimensionless conductance $g_0= 2\pi \hbar /e^2 R \gg
1$, where $R$ is the resistance of the contact, the amplitude of Coulomb
blockade oscillations in the ground state energy is exponentially
small,~\cite{Schon90} $\propto E_C \exp( -\frac{g_0}{2})$. The magnitude
and temperature dependence of Coulomb blockade oscillations in this
problem were studied in many theoretical works using the renormalization
group (RG) treatment~\cite{Falci,Hofstetter} and the
instanton~\cite{Korshunov87} approach~\cite{Zaikin,Grabert96} to the
dissipative action of Ref.~\onlinecite{Ambegaokar82}.

In the instanton approach the cost of one instanton is $\propto \exp
(-\frac{g(T)}{2})$, where $g(T) \approx g_0 -2 \ln\frac{g_0 E_C}{2\pi^2
  T}$,~\cite{Falci} is the renormalized conductance.  The instanton
gas can be considered as non-interacting at relatively high temperatures,
when the renormalized conductance $g(T)$ is still large.  In this regime
the thermodynamic potential depends sinusoidally on the gate
voltage,~\cite{Grabert96,Kamenev02}
\begin{equation}
\label{eq:chargingenergy}
        \Omega_{osc}(q)= - \tilde E_C(T) \cos(2\pi q),
\end{equation}
where $\tilde E_C(T)$ is the renormalized temperature-dependent charging
energy.  

Integration over the massive fluctuations around the instanton in
the Gaussian approximation leads to the logarithmic temperature
dependence~\cite{Grabert96} of the renormalized charging energy at
relatively high temperature. This results in the estimate $E_C g_0^3
\exp(-\frac{g_0}{2})$ for the amplitude of the Coulomb blockade
oscillations in ground state energy.

In this paper we evaluate non-Gaussian corrections for the fluctuations
about the instanton configurations in the lowest order of perturbation
theory in $1/g_0$. We show that these corrections diverge at $T\to 0$ and
significantly modify the magnitude and the temperature dependence of the
preexponential factor of the Coulomb blockade oscillations even at
relatively high temperatures, where the renormalized conductance $g(T)$ is
large and non-interacting instanton gas approximation is still valid. We
then apply a two-loop renormalization group (RG) to determine the
preexponential factor in the renormalized charging energy beyond the
region of applicability of perturbation theory.

The main result of this paper is the following expression for the
renormalized charging energy in Eq.~(\ref{eq:chargingenergy}),
\begin{subequations}
\label{results}
\begin{equation}
\label{result4}
\tilde E_C(T)
=\frac{E_C g_0^{5/2} }{3\pi^2 g(T)}
\left[g^{3/2}(T_0)-g^{3/2}(T)\right] e^{-g_0/2}. 
\end{equation}
Here $g(T)$ is the renormalized temperature dependent conductance, and
$T_0=\frac{E_C}{2\pi^2}$. With sufficient
accuracy $g(T_0)$ can be found using the perturbation theory,
$g(T_0)=g_0-2\ln g_0$. At lower temperatures,
but still such that $g(T) \gg 1$ the renormalized conductance $g(T)$ can
be found from the implicit relation,
\begin{equation}
\label{result5}
g(T)=g_0-2\ln\frac{g_0 T_0}{ T}+2\ln\frac{g(T)}{g_0},
\end{equation}
\end{subequations} 
which solves the two-loop renormalization group
equation.~\cite{Hofstetter}

Equations (\ref{eq:chargingenergy}) and (\ref{results}) rely on the
non-interacting instanton gas approximation and are valid at $T >
E_Cg_0^4 e^{-g_0/2}$.  Assuming that at lower temperatures the
amplitude of Coulomb blockade oscillations is only weakly
temperature-dependent we obtain the estimate for the magnitude of
oscillations in the ground state energy, $E_Cg_0^4 e^{-g_0/2}$.  This
is by a factor of $g_0$ greater than the result of
Ref.~\onlinecite{Grabert96}.

The paper is organized as follows: In Sec.~\ref{model} we describe the
dissipative action approach to the problem. In
Sec.~\ref{sec:single_instanton} we discuss the instanton gas
approximation. In Sec.~\ref{sec:gauss} we treat the fluctuations around
the instantons in the Gaussian approximation, and in
Sec.~\ref{sec:non-gauss} we obtain the leading $1/g_0$ correction to the
Gaussian result.  In Sec.~\ref{sec:RG} we use the two-loop renormalization
group to extend the perturbative results of Sec.~\ref{sec:non-gauss} to
the low temperature regime $\ln(E_C/T)\sim g_0$. Our results are
summarized in Sec.~\ref{conclusion}.

\section{\label{model} partition function}

Following Ref.~\onlinecite{Ambegaokar82} the partition function of the
system can be written as an imaginary time functional integral over the
auxiliary field $\phi(\tau)$ which decouples the Coulomb interaction
Eq.~(\ref{interaction}).  Different configurations of the field
$\phi(\tau)$ fall into different topological classes labeled by the
winding number $W$, characterizing the boundary conditions in imaginary
time: $\phi_{W}(\tau+\beta)=\phi_W(\tau) +2\pi W$, where $\beta=1/T$.  The
partition function is then written as a sum over the winding numbers,
\begin{equation}
\label{winding}
Z(q)=\sum\limits_{W=-\infty}^{+\infty}e^{2\pi iqW}Z_W,
\end{equation}
where $Z_W$ is a functional over the fields $\phi_W$ in the topological
class $W$ given by
\begin{equation}
\label{Zfunction}
Z_W=\int D[\phi_W]e^{-S[\phi_W]}.
\end{equation}
The action $S[\phi]$ has the following form,
\begin{equation}
\label{action1}
S[\phi]=S_d[\phi]+\int^{\beta}_{0} \frac{\dot{\phi}^2(\tau)}{4E_C}d\tau.
\end{equation}
The second term in the right hand side of Eq.~(\ref{action1}) is the
charging energy term.  The first term describing the tunnel junction was
obtained in Ref.~\onlinecite{Ambegaokar82} and can be conveniently written
using the complex variable
$u=\exp(2\pi i T\tau)$ as follows,
\begin{equation}
\label{action3}
S_d[\phi]=-g_0\oint \frac{du du_1}{(2\pi i)^2}
\frac{\rm Re\left(1-e^{i\phi(u)-i\phi(u_1)}\right)}{(u-u_1)^2}.
\end{equation}
The integral in this equation is taken over the unit circle, $|u|,|u_1|=1$.

The minimum of the dissipative action (\ref{action3}) in the topological
sector $W$ is equal to $g_0 |W|/2$. In the trivial topological sector,
$W=0$, it is achieved when $\phi(\tau)=0$. In the sectors $W=\pm 1$ it is
achieved on the instanton configurations~\cite{Korshunov87} which can be
written in the complex notations as~\cite{Nazarov99}
\begin{equation}
\label{instanton}
\exp(i\phi_z)=f(u)=\frac{u-z}{1-uz^*},
\end{equation}
where $z$ (with $|z|<1$ for $W=1$ and $|z|>1$ for $W=-1$) is a complex
number characterizing the position and the width of the instanton. In the
topological sector with $|W|>1$ the dissipative action (\ref{action3}) is
minimized on the multi-instanton configuration of the field $\phi$ given
by the product of $|W|$ single-instanton terms, as in the right hand side
of Eq.~(\ref{instanton}).

The topological sector with $W=0$ in Eq.~(\ref{winding}) does not
contribute to the oscillatory part of the thermodynamic potential, $
\Omega(q)=-T\ln Z(q)$.  At relatively high temperatures, when the
renormalized conductance is large, $g \gg 1$, the main non-zero
contribution to the oscillatory part of $\Omega(q)$ comes from the terms
with winding numbers $W=\pm1$ in Eq.~(\ref{winding}). All other terms in
Eq.~(\ref{winding}) are exponentially small in $g$, and the renormalized
charging energy in Eq.~(\ref{eq:chargingenergy}) can be expressed as
\begin{equation}
  \label{eq:omega}
  \tilde E_C(T)=2T\frac{Z_1}{Z_0}.
\end{equation}
We therefore concentrate below on the topological sector $W=1$.

\section{\label{sec:single_instanton} single instanton approximation}

Since the value of the dissipative action on the instanton configuration
$\phi_z$ in Eq.~(\ref{instanton}) is independent of the instanton
parameter $z$ it is convenient to write the fields in the topological
sector $W=1$ in the form
\begin{equation}
\label{phi}
\phi_1=\phi_z + \tilde \phi_z,
\end{equation}
where $\tilde \phi_z$ are massive fluctuations which are orthogonal to the
two zero modes of the dissipative action, $\partial \phi_z/\partial z$ and
$\partial \phi_z/\partial z^*$.  The renormalized charging energy
(\ref{eq:omega}) can be written as
\begin{equation}
  \label{eq:omegaz}
  \tilde E_C (T)=2T \int \frac{d^2z}{1-|z|^2}  Z_1(z),
\end{equation}
where $Z_1(z)$ is given by the following ratio of functional integrals
over the massive modes only,
\begin{equation}
  \label{eq:Zz}
  Z_1(z)=\frac{\int D [\tilde\phi_z](1-|z|^2) J(z,\tilde \phi_z) 
\exp(-S[\phi_z  +\tilde{\phi}_z] 
  )}{\int D [\phi_0] \exp(-S[\phi_0] )} ,
\end{equation}
where $J (z, \tilde\phi_z)$ the Jacobian of the variable transformation,
Eq.~(\ref{phi}).

Below we compute the renormalized charging energy $\tilde E_C$,
Eq.~(\ref{eq:omegaz}). In section \ref{sec:gauss} we evaluate the 
functional integral over the massive modes (\ref{eq:Zz}) in the Gaussian
approximation. In section \ref{sec:non-gauss} we obtain corrections to the
Gaussian approximation in leading order of perturbation theory in
$1/g_0$. Then, in section \ref{sec:RG} we employ the renormalization
group to obtain $\tilde E_C$ beyond the regime of validity of
perturbation theory.

\subsection{\label{sec:gauss}Gaussian approximation}

To leading order in $1/g_0$ we may evaluate the ratio of the functional
integrals in Eq.~(\ref{eq:Zz}) in the Gaussian approximation. To this end
we expand the actions in the numerator and in the denominator in
Eq.~(\ref{eq:Zz}) to second order in $\tilde \phi$ and $\phi_0$
respectively.

In the Matsubara basis,
\begin{equation}
\label{eq:matsubarabasis}
\phi_0 =\sum\limits_{n=-\infty}^{\infty} {\varphi}_n u^n,
\end{equation}
the action in the denominator acquires the following diagonal form,
\begin{equation}
\label{eq:gaussmatsubara}
S^0_{ \phi^2}= g_0 \sum\limits_{n\geq
  1}^{\infty}(n + a n^2)|\varphi_n|^2 , 
\end{equation}
where  we introduced the notation $a=\frac{2\pi^2 T}{g_0 E_C}$.

To evaluate the functional integral in the numerator in Eq.~(\ref{eq:Zz})
it is convenient to expand the massive fluctuations $\tilde{\phi}_z$ using
the basis of Ref.~\onlinecite{Korshunov87} which in the complex
notations~\cite{Nazarov99} can be written as, 
\begin{equation}
\label{basis}
\tilde \phi_z =\sum\limits_{n>0}\tilde \varphi_n u^n
f(u)+\sum\limits_{n<0}\tilde \varphi_n u^n f^*(u). 
\end{equation}
Here $f(u)$ is defined in Eq.~(\ref{instanton}).  In this basis to second
order in $\tilde{\phi}_z$ the dissipative part of the action,
Eq.~(\ref{action3}), has the following diagonal form,
\begin{equation}
\label{bare}
S^i_{\tilde \phi^2}= \frac{g_0}{2} + g_0 \sum\limits_{n\geq 1}^{\infty}n
|\tilde \varphi_n|^2.  
\end{equation}
The superscript $i$ here refers to the presence of the instanton, whereas
the superscript $0$ in Eq.~(\ref{eq:gaussmatsubara}) denotes the trivial
topological sector, $W=0$.

Instead of calculating the Jacobian $J(z, \tilde \phi_z)$ in
Eq.~(\ref{eq:Zz}) directly, we can express the integration measure through
the metric tensor $\hat{A}(z,\tilde{\phi}_z)$ using the
identity~\cite{Polyakov}
\begin{equation}
  \label{eq:intmeasure}
  J(z, \tilde \phi_z)D[\tilde{ \phi}_z]=\sqrt{ \rm{det}
  \hat{A}(z,\tilde{\phi}_z)} \prod 
  \limits_{n} 
  d\tilde{\varphi}_n .
\end{equation}
The metric tensor $\hat{A}(z,\tilde{\phi}_z)$ is presented in Appendix
\ref{Jacobian}. Fortunately we do not need to evaluate its determinant for
an arbitrary configuration of the field $\tilde{\phi}_z$. Indeed, the
fluctuations of the massive modes, $\tilde{\varphi}_n$ are small as
$1/g_0$, see Eq.~(\ref{bare}).  Therefore, for large $g_0$ we can expand
the determinant of the metric tensor in the powers of $\tilde{\varphi}_n$.
In leading order in $1/g_0$, i.e. in the Gaussian approximation it is
sufficient to evaluate $\rm{det} \hat{A}(z,\tilde{\phi}_z)$ on the
instanton configuration, $\tilde{\varphi}_n=0$. We show in Appendix
\ref{Jacobian} that
\begin{equation}
  \label{eq:Jacobian_0}
J(z,0)=  \sqrt{ \rm{det} \hat{A}(z, 0)}=\frac{1}{1-|z|^2}.
\end{equation}

The quadratic form of the charging part of the action given by the second
term in Eq.~(\ref{action1}) is not diagonal in the basis (\ref{basis}).
However, for the purpose of evaluating the functional integral in the
numerator of Eq.~(\ref{eq:Zz}) one can neglect its off-diagonal elements. 
We show in Appendix \ref{charging} that their contribution is small as
$1/g_0$.  The diagonal part of the charging action is
given by Eq.~(\ref{diagcharge}).  
Then, the ratio of the functional integrals
in (\ref{eq:Zz}) reduces to the product
\begin{equation}
\label{ratio1}
 \frac{g_0}{\pi} \prod_{n=2}^{\infty}\frac{n + a n^2}{n-1 + a n^2
  +\frac{2a|z|^2}{1-|z|^2} }= \frac{g_0}{\pi} a^{-1+\frac{2a|z|^2}{1-|z|^2}},
\end{equation}
where $a$ was defined below Eq.~(\ref{eq:gaussmatsubara}).  Since the
characteristic instanton frequencies are $T/(1-|z|^2) \leq E_C \ll g_0
E_C$ the second term in the exponent of this expression can be neglected.
Using the expression Eq.~(\ref{charginst}) for the charging action on
the instanton configuration we obtain in the Gaussian approximation,
\begin{equation}
  \label{eq:Zzgrabert}
Z^\textrm{G}_1(z)= 
\frac{g_0^2E_C}{2\pi^3 T}  \exp\left( -\frac{g_0}{2} -
  \frac{\pi^2 T[1+|z|^2]}{E_C[1-|z|^2]}\right). 
\end{equation}
Substituting this expression into Eq.~(\ref{eq:omegaz}) we obtain with
logarithmic accuracy,
\begin{equation}
  \label{eq:omegagrabert}
  \tilde E_C (T) = \frac{g_0^2E_C}{\pi^2}\ln\left[\frac{T_0}{
  T}\right] 
  \exp\left(-\frac{g_0}{2}\right),  
\end{equation}
where we introduced the notation
\begin{equation}
  \label{eq:T_0def}
  T_0=\frac{E_C}{2\pi^2}.
\end{equation}
Equation (\ref{eq:omegagrabert}) coincides with the result of
Ref.~\onlinecite{Grabert96}.  One can neglect the interaction between
instantons and obtain Eq.~(\ref{eq:omegagrabert}) only when $Z_1(z) \ll
1$.  In section \ref{sec:non-gauss} we show that even in this regime the
non-Gaussian corrections to the functional integral in Eq.~(\ref{eq:Zz})
lead to large corrections to the preexponential factor in
Eq.~(\ref{eq:omegagrabert}).

\subsection{Corrections to the Gaussian approximation}
\label{sec:non-gauss}

The Gaussian approximation discussed in section \ref{sec:gauss} is
asymptotically exact at $g_0 \to \infty$. The corrections to it are small
in $1/g_0$ and may be evaluated perturbatively.  To obtain the leading
$1/g_0$ correction it is sufficient to expand the Jacobian in
Eq.~(\ref{eq:Zz}) to second order in $\tilde \phi_z$ and to expand the
actions in the numerator and in the denominator in Eq.~(\ref{eq:Zz}) to
fourth order in $\tilde \phi_z$ and $\phi_0$ respectively:
\begin{subequations}
\label{eq:expansionab}
\begin{eqnarray}
\label{expansioni}
S_d[\phi_1]&=&S^i_{\tilde \phi^2}+S^i_{\tilde \phi^3}+S^i_{\tilde
  \phi^4}, \\
 \label{expansion0}
S_d[\phi_0]&=&S^0_{\phi^2}+S^0_{ \phi^4}.
\end{eqnarray}
\end{subequations}
The terms $S^0_{\phi^2}$ and $S^i_{\tilde \phi^2}$ were defined in
Eqs.~(\ref{eq:gaussmatsubara}) and (\ref{bare}) respectively, and
$S^i_{\tilde \phi^3}$, $S^i_{\tilde \phi^4}$, and $S^0_{ \phi^4}$
are given by the following equations;
\begin{subequations}
\label{expansion2}
\begin{eqnarray}
S_{\tilde \phi^4}^i &=&
\frac{g_0}{24}\oint\frac{dudu_1}{(2 \pi i)^2} 
{\rm Re}\left[\frac{f(u)}{f(u_1)}\right] 
\frac{ (\tilde \phi- \tilde
    \phi_1)^4}{(u-u_1)^2}, 
\\  S^i_{\tilde \phi^3} &=&
\frac{g_0}{6}\oint\frac{dudu_1}{(2\pi)^2} {\rm
  Im}\left[\frac{f(u)}{f(u_1)}\right]\frac{(\tilde \phi-\tilde  \phi_1)^3 
}{(u-u_1)^2}   , \\ 
 S_{ \phi^4}^0  &=&
\frac{g_0}{24}\oint\frac{dudu_1}{(2 \pi i)^2} 
\frac{ (\phi-  \phi_1)^4}{(u-u_1)^2}.
\end{eqnarray}
\end{subequations}
Here the function $f(u_i)$ was defined in Eq.~(\ref{instanton}), and we
introduced the short hand notations $\tilde \phi_i=\tilde \phi(u_i)$.
Substituting Eqs.~(\ref{eq:expansionab}) into Eq.~(\ref{eq:Zz}) and using
Eqs.~(\ref{eq:intmeasure}) and (\ref{eq:Jacobian_0}) we obtain up to terms
of order $1/g_0$,
\begin{eqnarray}
\frac{Z_1(z)}{Z^\textrm{G}_1(z)} &=& (1-|z|^2)\left
    \langle \sqrt{\textrm{det} \hat{A}(z,\tilde \phi_z)} \right \rangle_i
    \nonumber \\ 
&& -\left
    \langle S_{\tilde \phi^4}^{i} \right \rangle_i +
\left \langle S_{ \phi^4}^0\right \rangle_0 +\frac{1}{2} 
\left \langle (S^i_{\tilde \phi^3})^2\right \rangle_i ,
\label{expansion}
\end{eqnarray}
where $\langle \ldots\rangle_i$ and $\langle \ldots\rangle _0$ denote
averaging with respect to the Gaussian actions in the presence and in the
absence of the instanton respectively.

\begin{figure}
\resizebox{.38\textwidth}{!}{\includegraphics{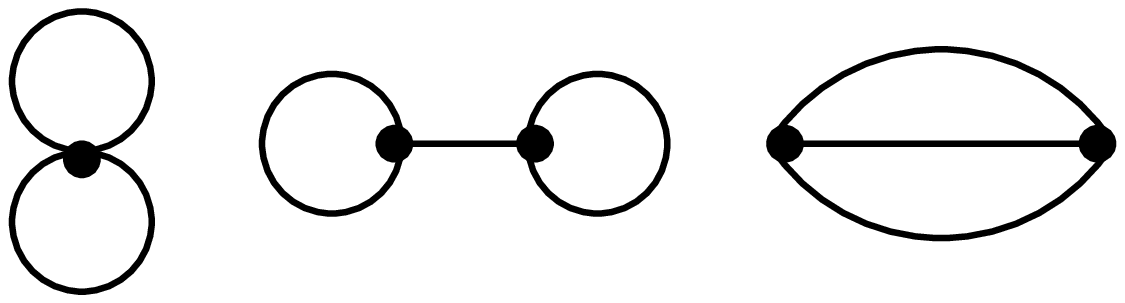}}
\caption{\label{diagrams}  Diagrams  representing 
  $1/g_0$ corrections to $\frac{Z_1(z)}{Z_1^G(z)}$ in
  Eq.~(\ref{expansion}). The vertices are proportional to $g_0$, and the
  Gaussian propagators of fields $\phi$ depicted by the solid lines are
  proportional to $1/g_0$. }
\end{figure} 

The first term in the right hand side of Eq.~(\ref{expansion}) needs to be
evaluated to second order in $\tilde \phi_z$. This is carried out in
Appendix \ref{Jacobian}.  The result is given by Eq.~(\ref{finalJ}).  The
calculation of the other terms is facilitated by the use of Wick's theorem
and reduces to evaluating the diagrams in Fig.~\ref{diagrams}. The
calculations are straightforward and are presented in Appendix \ref{A}.
Only fluctuations of $\phi$ with relatively low frequencies, $\leq
T/(1-|z|^2)$, contribute to the corresponding functional integrals. We can
therefore neglect the charging energy term in the Gaussian action for the
massive modes.  The calculations are further simplified by observing that
the two-point correlation functions in the diagrams in Fig.~\ref{diagrams}
become diagonal in the bases (\ref{eq:matsubarabasis}) and (\ref{basis}).
We then obtain with logarithmic accuracy,
\begin{equation}
  \label{eq:Zzpert}
  Z_1(z) = Z^\textrm{G}_1(z)\left[1+\frac{1}{g_0}\ln
  \frac{g_0 T_0}{T}-\frac{1}{g_0}\ln(1-|z|^2) \right], 
\end{equation}
where $T_0$ was defined in Eq.~(\ref{eq:T_0def}).  Substituting
Eqs.~(\ref{eq:Zzpert}) and (\ref{eq:Zzgrabert}) into (\ref{eq:omegaz}) and
using Eq.~(\ref{finalJ}) we obtain for the renormalized charging energy,
\begin{equation}
\label{result1}
\tilde E_C (T)= \frac{g_0^2E_C}{\pi^2}\ln\frac{T_0}{ T}
\left[1+ \frac{\ln \frac{g_0 T_0}{ T}}{g_0}
+\frac{\ln\frac{T_0}{ T}}{2g_0}
\right]e^{-g_0/2}.
\end{equation}

Equations (\ref{eq:Zzpert}) and (\ref{result1}) represent the main results
of this section.  The first term in the right hand side of
Eq.~(\ref{result1}) coincides with the result of
Ref.~\onlinecite{Grabert96}, and the others represent a perturbative
correction arising from taking into account non-Gaussian fluctuations.
Equation (\ref{result1}) is valid at relatively high temperatures,
$\ln\frac{T_0}{ T}\ll g_0$, when the non-Gaussian correction is small. We
consider the region of lower temperatures, $\ln\frac{T_0}{ T} \sim g_0$,
in section \ref{sec:RG} using the renormalization group approach.

\section{\label{sec:RG} Renormalization Group}

At low temperatures, when $\ln\frac{g_0 T_0}{ T}$ becomes of the
order of $g_0$, the non-Gaussian correction in Eq.~(\ref{eq:Zzpert})
becomes significant, and the perturbative expressions (\ref{eq:Zzpert})
and (\ref{result1}) are no longer valid.  There is a wide temperature
range in which the perturbative approach used in section
\ref{sec:single_instanton} fails but the renormalized conductance is still
large, $g(T) \approx g_0 -2 \ln\frac{g_0 T_0}{ T} \gg 1$, and
therefore the single instanton approximation, Eq.~(\ref{eq:omegaz}) is
still valid.  Below we apply the renormalization group to study the
amplitude of the Coulomb blockade oscillations in this regime.

From Eq.~(\ref{eq:Zzgrabert}) we observe that $Z_1(z)$ depends
exponentially on the renormalized conductance, $Z_1(z)\propto
\exp\left(-g(T)/2\right)$.  Therefore to obtain the preexponential factor
in the renormalized charging energy in Eq.~(\ref{eq:omegaz}) it is
necessary to compute the renormalized conductance using the two-loop
renormalization group.

We obtain and solve the two-loop RG equations for the conductance in
Sec.~\ref{sec:RG_conductance}. Then, in Sec.~\ref{sec:RG_Z} we obtain the
renormalized charging energy $\tilde{E}_C$, Eq.~(\ref{eq:omegaz}), by
evaluating the functional integral $Z_1(z)$ in Eq.~(\ref{eq:Zz}) with the
aid of the renormalization group.

\subsection{\label{sec:RG_conductance} Renormalized conductance}

\begin{figure}
\resizebox{.38\textwidth}{!}{\includegraphics{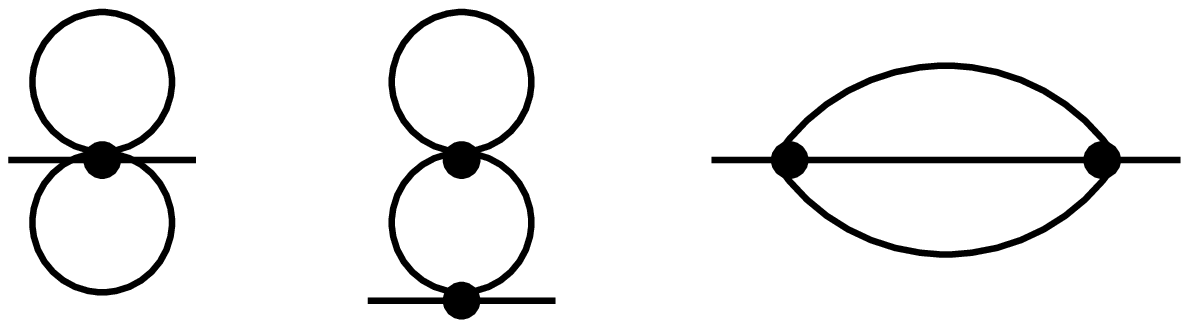}}
\caption{\label{fig:2}  Diagrams of order $1/g_0$ for the renormalized
  conductance $g(\nu)$.}   
\end{figure}

To obtain the two-loop RG equations we expand the dissipative action in
Eq.~(\ref{action3}) to sixth order in $\phi_0$.  Next we introduce a
running frequency scale $\nu$ and write $\phi_0 =\phi_0^s+ \phi_0^f$.
Here $\phi_0^f$ represents fast degrees of freedom with Matsubara
frequencies $\nu_n= T n$ satisfying the inequality $\nu <\nu_n < g_0T_0$,
and $\phi_0^s$ represents slow degrees of freedom with $ \nu_n < \nu $.
The calculation of the renormalized conductance at frequency $\omega$
amounts to evaluating the diagrams in Fig.~\ref{fig:2}, where internal
lines correspond to the propagators of the fast degrees of freedom.

The role of the charging term in the Gaussian action,
Eq.~(\ref{eq:gaussmatsubara}), amounts merely to the ultraviolet cutoff of
the frequency integrals over the internal lines at frequencies $g_0 T_0$,
where $T_0$ was defined in Eq.~(\ref{eq:T_0def}).  As a result we obtain,
\begin{equation}
\label{phi4}
g(\nu)=g_0-2\ln\frac{g_0T_0}{\nu}
-\frac{4}{g_0}\ln\frac{g_0T_0}{\nu}. 
\end{equation}
From Eq.~(\ref{phi4}) we obtain the following two-loop renormalization
group equation for the conductance in agreement with
Ref.~\onlinecite{Hofstetter},
\begin{equation}
\label{RG}
\frac{dg(\nu)}{d\xi}=-2-\frac{4}{g(\nu)}, \hspace{.6cm}  \xi = \ln
\frac{\Lambda}{\nu}, 
\end{equation}
where $\Lambda = g_0 T_0$ is an ultraviolet cutoff.  The first term
in the right hand side of Eq.~(\ref{RG}) reproduces the one-loop
renormalization group equation of Refs.~\onlinecite{Kosterlitz,Falci}.

By integrating the two-loop renormalization group equation, Eq.~(\ref{RG})
from $\nu= g_0 T_0$ to $\nu = T$ we obtain Eq.~(\ref{result5}) which
implicitly determines the temperature dependence of the renormalized
conductance $g(T)$.

\subsection{\label{sec:RG_Z} Renormalized charging energy}

Next, we apply the renormalization group to evaluate $Z_1(z)$ in
Eq.~(\ref{eq:Zz}).  We start by considering the case of the so-called line
instantons, $z=0$. According to the RG procedure the functional
integration is performed by separating $\phi$ into fast and slow degrees
of freedom, $\phi=\phi^f+\phi^s$, where $\phi^f$ includes fluctuations
with Matsubara frequencies $\nu_n$ in a narrow interval,
$\nu=\Lambda e^{-\xi}< \nu_n < \Lambda$ . Upon the integration over
the fast degrees of freedom the remaining functional integral acquires a
factor resulting from integrating out the fast fluctuations of $\phi$.
This factor depends on the running coupling constant $g(\nu)$. The
remaining action for the slow fluctuations is characterized by a
renormalized conductance. Due to this multiplicative nature of the
functional integration we can express the logarithmic derivative of $
Z_1(0)$ through a function of the renormalized conductance $g(\nu)$
only, $\frac{d \ln Z_1(0)}{d\xi}=f[g(\nu)]$. From the perturbative
results (\ref{eq:Zzpert}) and (\ref{eq:Zzgrabert}) we obtain for the
logarithmic derivative,
\begin{equation}
\label{Z10}
\frac{d \ln Z_1(0)}{d\xi}=1+\frac{1}{g(\nu)}.
\end{equation}
Using the two-loop renormalization group for the conductance,
Eq.~(\ref{RG}) we obtain,
\begin{equation}
  \label{eq:Z1RGequation}
 \frac{d \ln Z_1(0)}{d g(T)} = - \frac{1}{2}+\frac{1}{2g(T)}.
\end{equation}
The solution of this equation which satisfies the high temperature
asymptotic (\ref{eq:Zzpert}), (\ref{eq:Zzgrabert}) is given by,
\begin{equation}
  \label{eq:Z1RG}
  Z_1(0)= \frac{\sqrt{g_0g(T)}}{2\pi^3}\exp\left[-\frac{g(T)}{2}\right].
\end{equation}
One can directly check this expression against the perturbative result
Eq.~(\ref{eq:Zzpert}). To obtain $Z_1(0)$ to order $1/g_0$ we need to
evaluate $g(T)$ in the exponent of this equation to the two-loop order,
and $g(T)$ in the preexponential factor to the one-loop order only.

To evaluate $Z_1(z)$ at finite $z$ it is convenient to express it in the
form 
\begin{equation}
  \label{eq:Xdef}
  Z_1(z)=Z_1(0)X(z)\exp\left(-\frac{\pi^2
    T[1+|z|^2]}{E_C[1-|z|^2]}\right),
\end{equation}
where the charging energy on the
instanton configuration is written out explicitly, c.f.
Eq.~(\ref{eq:Zzgrabert}).

The perturbative result (\ref{eq:Zzpert}) implies $X(z)=1
-\frac{1}{g_0}\ln (1-|z|^2) $, where the logarithmic term arises from the
integration over the fluctuations of $\phi$ with frequencies ranging
between $T$ and the instanton frequency $\nu_z= T/(1-|z|^2)$,
see the discussions above Eq.~(\ref{eq:Zzpert}) and in Appendix \ref{A}.
Therefore in the renormalization group treatment the conductance $g_0$ in
the logarithmic correction should be understood as the renormalized
conductance $g(\nu_z)\approx g(T) +2 \ln\frac{\nu_z}{T}$ at the
instanton frequency $\nu_z$.  Using the considerations presented above
Eq.~(\ref{eq:Z1RG}) and the perturbative result, Eq.~(\ref{eq:Zzpert}) we
express the logarithmic derivative of $X(z)$ through the renormalized
conductance $g(\nu_z)$ as,
\begin{equation}
  \label{eq:XRG}
\frac{d\ln [X(z)]}{d\ln\nu_z}=\frac{1}{g(\nu_z)}.  
\end{equation}
Using the one-loop renormalization group equation $\frac{d
  g(\nu_z)}{d\ln \nu_z}=2$ and the boundary condition $X(0)=1$ we
find,
\begin{equation}
\label{Xresult}
X(z)=\sqrt{\frac{g(\nu_z)}{g(T)}}.
\end{equation}
Substituting Eq.~(\ref{eq:Xdef}) into Eq.~(\ref{eq:omegaz}) we write the
renormalized charging energy as
\[
\tilde E_C (T) = 2T Z_1(0)\int
\frac{d^2z}{1-|z|^2} X(z)\exp\left[-\frac{\pi^2
T[1+|z|^2]}{E_C[1-|z|^2]}\right] .   
\]
Next, we express the integration measure through the differential of the
renormalized conductance, $dg(\nu_z)$, according to the one-loop RG
equation, $\frac{d^2z}{1-|z|^2} = -\pi d \ln (1-|z|^2) = \frac{\pi}{2} d
g(\nu_z)$. The resulting integral over the $ d g(\omega_z)$ ranges from
$g(\omega_z)=g(T)$ to $g(\nu_z)=g(T_0)=g_0-2\ln g_0$. Using
Eqs.~(\ref{eq:Z1RG}) and (\ref{Xresult}) we obtain for the renormalized
charging energy,
\begin{equation}
\label{results6}
\tilde E_C(T)
=\frac{T \sqrt{g_{0}}}{3\pi^2} 
\left[g^{3/2}(T_0)-g^{3/2}(T)\right] e^{-g(T)/2}, 
\end{equation}
where $T_0$ was defined in Eq.~(\ref{eq:T_0def}).

Next we express the temperature $T$ through the renormalized conductance
$g(T)$ using the solution of the two-loop RG equation,
Eq.~(\ref{result5}).  Substituting this result into Eq.~(\ref{results6})
we obtain Eq.~(\ref{result4}).

Equations (\ref{results}) represent the main result of this paper. They
determine the temperature dependence of the amplitude of Coulomb blockade
oscillations in a wide temperature regime where the renormalized
conductance $g(T)$ is large and the non-interacting instanton gas
approximation~\cite{Grabert96} is valid.

At relatively high temperatures, when $\ln \frac{g_0 T_0}{ T}\ll
g_0$, Eq.~(\ref{results}) reproduces the perturbative result,
Eq.~(\ref{result1}).  At lower temperatures, when $\ln \frac{g_0 T_0}{T}
\sim g_0$, the renormalized charging energy in Eq.~(\ref{result4})
significantly exceeds the prediction of the Gaussian
approximation~\cite{Grabert96}, Eq.~(\ref{eq:Zzgrabert}).

The non-interacting instanton gas approximation is valid while $Z_1/Z_0 <
1$.  Let us extrapolate the results (\ref{results}) to the low temperature
limit, when $Z_1/Z_0 \sim 1$. This happens at $T \sim E_C
g_0^4 e^{-g_0/2}$ when $g(T)\approx 4\ln g_0$. Substituting this
conductance into Eq.~(\ref{result4}) we obtain the following estimate for
the amplitude of Coulomb blockade oscillations in the ground state energy,
$\tilde E_C \sim E_C g_0^4 e^{-g_0/2}$.  This exceeds the estimate
arising from the Gaussian treatment of the fluctuations~\cite{Grabert96}
by a large factor, $g_0$.

\section{\label{conclusion} discussion}

\subsection{\label{sec:summary} Summary of the results}

We studied the amplitude, $\tilde E_C (T)$, of the Coulomb blockade
oscillations in the thermodynamic potential for a metallic grain connected
to a lead by a tunneling contact with a large tunneling conductance $g_0$.
The effects of a finite mean level spacing in the
grain~\cite{Beloborodov02} were neglected. We worked withing the
non-interacting instanton gas approximation.  By applying perturbation
theory we obtained the leading $1/g_0$ correction to the Gaussian
result~\cite{Grabert96} for the renormalized charging energy,
Eq.~(\ref{result1}).  Combining the instanton analysis with the two-loop
renormalization group we found the temperature dependence of the
renormalized charging energy $\tilde E_C(T)$, and of the renormalized
conductance $g(T)$, Eq.~(\ref{results}), in a wide temperature range $E_C
g_0^4 e^{-g_0/2}<T<E_C$. The use of the two-loop RG enables us to
determine the preexponential factor in the temperature dependence of
$\tilde E_C(T)$ even at relatively low temperatures, when $\ln\frac{g_0
  T_0}{ T} \sim g_0$. In this regime the renormalized charging
energy in Eq.~(\ref{result4}) is significantly greater than the result of
the Gaussian approximation, Eq.~(\ref{eq:omegagrabert}).

Assuming that the amplitude of the Coulomb blockade oscillations weakly
depends on temperature at $T < E_Cg_0^4 e^{-g_0/2}$ from
Eq.~(\ref{results}) we obtain the following estimate for the amplitude of
the oscillations in the ground state energy, $\tilde E_C \sim E_C g_0^4
e^{-g_0/2}$.  This estimate is greater than the result of the Gaussian
approximation~\cite{Grabert96} by a factor of $g_0$.  In
Ref.~\onlinecite{Zaikin} the result for the integral over the massive
fluctuations around the instanton configuration differed from that in
Ref.~\onlinecite{Grabert96} and in Eq.~(\ref{eq:Zzgrabert}). This lead to
a different estimate for the amplitude of the Coulomb blockade
oscillations in the ground state energy, $ E_C g_0^2e^{-g_0/2}$.  In
Ref.~\onlinecite{Hofstetter} the instanton effects were ignored, and the
estimate $E_C g_0e^{-g_0/2}$ for the amplitude of the ground state energy
oscillations was obtained using the two-loop renormalization group.  The
topological effects, on the other hand, may drastically alter the results
of a perturbative RG consideration.  This is well known, for example, in
the theory of antiferromagnetic Heisenberg spin chains,~\cite{Fradkinbook}
where the perturbative RG predicts the appearance of a spin gap for an
arbitrary value of the spin $S$. However, for half-integer spin the
topological effects are expected to lead to a gapless
state.~\cite{Haldane}

\subsection{\label{sec:numerics} Applicability of the results and comparison
  with numerical studies}

Within the framework of the model considered here the thermodynamics of
the grain is described by Eqs.~(\ref{winding})-(\ref{action3}).  The
partition function, Eq.~(\ref{winding}), depends on two dimensionless
parameters, $g_0$ and $T/E_C$.  This model has been studied numerically by
several groups.~\cite{Herrero,Wang}

To compare our results with the numerical studies it is important to keep
in mind the approximations that were made. These approximations determine
the region of applicability of the instanton approach in the space of
parameters $g_0$ and $T/E_C$.  Below we review the approximations made: 

i) The instanton configuration was found by neglecting the charging part
of the action in comparison with the dissipative part. This is justified
if $a=\frac{2\pi^2T}{g_0 E_C}=\frac{T}{g_0T_0} \ll 1$.  

ii) To obtain the Gaussian result, Eq.~(\ref{eq:omegagrabert}), we assumed
that the charging action on the instanton configuration, second term in
the exponent in Eq.~(\ref{eq:Zzgrabert}), is negligible for long
instantons, $z\to 0$.  This is valid for $T \ll T_0$.  In addition,
Eq.~(\ref{eq:omegagrabert}) holds with logarithmic accuracy. This implies
not only that $\frac{T_0}{ T} \gg 1$ but that $\ln\frac{T_0}{ T} \gg 1$.
The low temperature regime is difficult to study numerically. At the
lowest available temperature in Ref.~\onlinecite{Herrero},
$\ln\frac{T_0}{T} =3.2$.  Thus, even at the lowest temperature, the result
of Ref.~\onlinecite{Zaikin} is not parametrically different from that of
Ref.~\onlinecite{Grabert96}, Eq.~(\ref{eq:omegagrabert}).

iii) The use of the single instanton approximation imposes the lower
bound, $T^*$, on the temperature range of applicability of the results.
In Ref.~\onlinecite{Kamenev02} the two-instanton contribution to the
partition function was taken into account in the one-loop approximation.
It was shown that the single instanton approximation,
Eq.~(\ref{eq:omega}), applies as long as $\frac{Z_1}{Z_0} \ll 1$, i.e. for
$T^* \gg \tilde{E}_C(T^*) $.  It is easy to see from
Eq.~(\ref{eq:omegagrabert}) that $T^*$ decreases as the dimensionless
conductance $g_0$ grows. The largest conductance for which the temperature
dependent data are available in Ref.~\onlinecite{Herrero} is $g_0=2\pi^2$.
Setting $\frac{\tilde{E}_C(T^*)}{T^*} =0.3$ and using
Eq.~(\ref{eq:omegagrabert}) we find that $\frac{T^*}{E_C}\approx 10^{-2}$.
The temperature dependence of $E^*_C=2\pi^2 \tilde{E}_C$ in
Ref.~\onlinecite{Herrero} saturates roughly at the same temperature.  At
this temperature $\ln\frac{T_0}{ T^*} \approx 1.6$ and quantitative
comparison of our results with numerical results is not justified.
Qualitatively however, at this temperature the results of
Eq.~(\ref{eq:omegagrabert}) and of Ref.~\onlinecite{Zaikin} are not very
different.

Thus quantitative comparison of numerical results with the results of
different analytical approaches, Refs.~\onlinecite{Zaikin,Grabert96} and
Eq.~(\ref{results}) requires numerical studies at much lower temperatures
and larger conductances than those available at present.

\begin{acknowledgments}
  
  We are especially grateful to A.~Kamenev and K.~Matveev for very useful
  discussions.  We also thank M.~Feigelman, F.~Hekking, E.~Mishchenko,
  M.~Skvortsov, A.~Tsvelik and A.~Vainstein for valuable discussions. We
  gratefully acknowledge the warm hospitality of the Institute for
  Theoretical Physics in Santa Barbara, where part of this work was
  performed. A.~A. and I.~B. were sponsored by the Grants DMR-9984002,
  BSF-9800338 and by the A.~P.~Sloan and the Packard Foundations. A.~L.
  was partially supported by the NSF grant N 0120702.

\end{acknowledgments}

\appendix

\section{\label{Jacobian} Evaluation of $\textrm{ det}
  \hat{A}(z,\tilde{\phi}_z)$ }

\begin{figure}
\resizebox{.38\textwidth}{!}{\includegraphics{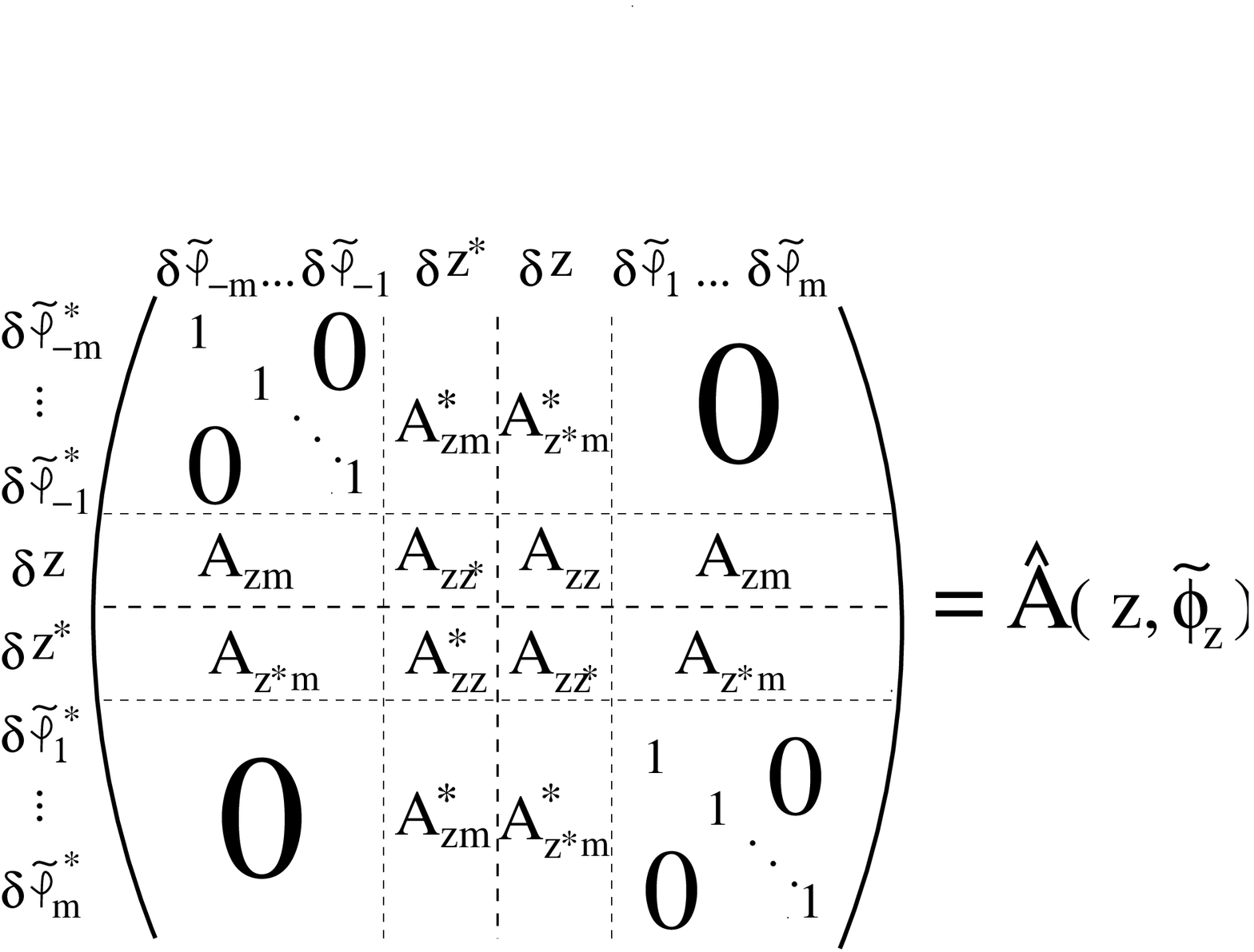}}
\caption{\label{matrix}  Matrix elements of the metric tensor
  $\hat{A}(z,\tilde{\phi}_z)$ in coordinates $z$, $\{ \tilde{\varphi}_n\}$. }
\end{figure}

To obtain the metric tensor $\hat{A}(z,\tilde{\phi}_z)$ in
Eq.~(\ref{eq:intmeasure}) we express the variation of the field $\phi$
through the variables $z$ and $\{ \tilde{\varphi}_n\}$ with the aid of
Eqs.~(\ref{phi}) and (\ref{basis}),
\begin{eqnarray}
\label{variation}
\delta \phi &=& \frac{\partial \phi_z}{\partial z}\delta z +
\sum\limits_{n>0}\tilde \varphi_n u^n \frac{\partial f}{\partial z} \delta
z + \sum\limits_{n<0} 
\tilde \varphi_n u^n \frac{\partial f^*}{\partial z} \delta z \nonumber \\ 
&+& 
\frac{\partial \phi_z}{\partial z^*}\delta z^* + \sum\limits_{n>0}\tilde
\varphi_n u^n \frac{\partial f}{\partial z^*} \delta z^* +
\sum\limits_{n<0} 
\tilde \varphi_n u^n \frac{\partial f^*}{\partial z^*} \delta z^*
\nonumber \\ 
&+& \sum\limits_{n>0}\delta \tilde \varphi_n u^n
f(u)+\sum\limits_{n<0}\delta \tilde \varphi_n u^n f^*(u).
\end{eqnarray}
Below we omit the arguments $z$ and $\tilde{\phi}_z$ of the metric tensor
$\hat{A}$.  Its matrix elements are obtained from the following relation,
\begin{eqnarray}
\oint\frac{du}{2\pi i u}|\delta
\phi|^2 &=& \sum\limits_{n,m}
A_{nm}\delta\tilde{\varphi}^*_n
\delta\tilde{\varphi}_m  + 2A_{zz^*}\delta z \delta z^* \nonumber \\
&+&A_{zz}^*\delta z^* \delta z^* + A_{zz}\delta z\delta z
\nonumber\\  
&+&2 \sum\limits_{m} \left[ A_{zm}\delta z +
  A_{z^*m}\delta z^* \right] 
\delta\tilde{\varphi}_m . \label{measure}
\end{eqnarray} 
Substituting Eq.~(\ref{variation}) into Eq.~(\ref{measure}) it is
straightforward to find all the elements of the metric tensor
$\hat{A}(z,\tilde{\phi}_z)$. Most of the matrix elements of
$\hat{A}(z,\tilde{\phi}_z)$ vanish.  The schematic form of this matrix is
shown in Fig.~\ref{matrix}.  

The determinant of $\hat{A}(z,\tilde{\phi}_z)$ in Eq.~(\ref{eq:intmeasure})
needs to be evaluated up to terms of order $1/g_0$.  Therefore it is
sufficient to find the matrix element $A_{z z}$ up to the linear order in
$\tilde{\varphi}$. Integrating over the variable $u$ in
Eq.~(\ref{measure}) we obtain the following expressions for the elements
of matrix $\hat{A}(z,\tilde{\phi}_z)$,
\begin{subequations}
\label{elements}
\begin{equation}
\label{elements_a}
A_{zz} = -i \sum\limits_{n>0}\tilde{\varphi}_n\frac{z^{n-1}}{1-|z|^2} - 
i \sum\limits_{n<0}\tilde{\varphi}_n^*\frac{z^{-n-1}}{1-|z|^2}, 
\end{equation}
\begin{equation}
  \label{elements_b}
  A_{zz^*} = \frac{1}{1-|z|^2}\left(1+
  2\sum\limits_{n>0}|\tilde{\varphi}_n|^2\right), 
\end{equation}
\begin{equation}
  \label{elements_c}
  A_{zm} = \left\{ 
\begin{array}{ll} 
\sum\limits_{n>0}\tilde{\varphi}_n^* z^{m-n-1}\theta(m-n),& m>0, \\
 - \sum\limits_{n<0}\tilde{\varphi}_n^*
z^{m-n-1}\theta(m-n), & m<0, 
\end{array} \right.
\end{equation}
\label{elements_d}
\begin{equation}
A_{z^* m} =\left\{ 
\begin{array}{ll} 
- \sum\limits_{n>0}\tilde{\varphi}_n^* (z^*)^{n-m-1}\theta(n-m),&  m>0, \\ 
\sum\limits_{n<0}\tilde{\varphi}_n^* (z^*)^{n-m-1} \theta(n-m),&  m<0 .
\end{array} \right.
\end{equation}
\end{subequations}
Here the $\theta$-function should be understood as $\theta(0)=0$.  We note
that even for very long instanton, $z \rightarrow 0$, some off-diagonal
elements of matrix $\hat{A}(z,\tilde{\phi}_z)$ remain finite, see
Eq.~(\ref{elements}).

To evaluate the renormalized charging energy $\tilde{E}_C$ in the Gaussian
approximation, see Sec.~\ref{sec:gauss}, the Jacobian in
Eq.~(\ref{eq:intmeasure}) may be evaluated on the instanton trajectory,
$\tilde{\phi}_z=0$. In this case the determinant of
$\hat{A}(z,\tilde{\phi}_z)$ is readily evaluated, and we obtain
Eq.~(\ref{eq:Jacobian_0}).

In order to obtain the leading $1/g_0$ correction to the Gaussian result
for $\tilde{E}_C$, we need to calculate the square root of the determinant
of $\hat{A}(z,\tilde{\phi}_z)$ in Eq.~(\ref{expansion}) to second order
in $\tilde{\phi}_z$.  To this end we expand
$\textrm{det}\hat{A}(z,\tilde{\phi})$ to second order in its off-diagonal
elements and perform the averaging with respect to the Gaussian
fluctuations around the instanton, $\left\langle \ldots\right\rangle_i$,
in Eq.~(\ref{expansion}).  We show in Appendix \ref{charging} that the
off-diagonal elements of the charging part of the action may be neglected,
and therefore $\left\langle \tilde{\varphi}_n \tilde{\varphi}_m^*
\right\rangle_i = \delta_{nm}\left(g_0
  |n|+\frac{2\pi^2T}{E_C}\left[(|n|+1)^2+\frac{2|z|^2}{1-
      |z|^2}\right]\right)^{-1}$.  Using this result and
Eq.~(\ref{elements}) we obtain,
\begin{equation}
\label{finalJ}
\left\langle \sqrt{ \textrm{det}\hat{A}(z,\tilde{\phi})}\right\rangle_i =
\frac{1}{1-|z|^2}\left(1 + \frac{1}{g_0}\ln \frac{g_0 T_0}{T}\right). 
\end{equation}

\section{\label{charging} Charging part of the action}

To obtain the charging energy, Eq.~(\ref{result1}) we evaluated the
functional integral over the massive fluctuations about the instanton in
the basis of Eq.~(\ref{basis}). We neglected the off-diagonal elements in
the charging part of the action. In this appendix we show that this
approximation is justified. Taking the off-diagonal matrix elements into
account leads to a correction to $\tilde{E}_C$ which is smaller than the
non-Gaussian correction given by the last two terms in Eq.~(\ref{result1})
by a factor of $\ln\frac{E_C}{T}$.  Therefore it can be neglected at low
temperatures.

To demonstrate this we rewrite the charging part of the action using the
complex variable $u=\exp(2\pi i T\tau)$ and Eq.~(\ref{phi}),
\begin{eqnarray}
\label{charging1}
&&\int^{\beta}_{0} \frac{\dot{\phi}^2(\tau)}{4E_C}d\tau = 
\frac{\pi i T}{2E_C}\oint u du   \nonumber \\
&& \times \left[\left(\frac{\partial\phi_z}{\partial
      u}\right)^2+2\frac{\partial\phi_z}{\partial
    u}\frac{\partial\tilde{\phi}_z}{\partial
    u}+\left(\frac{\partial\tilde{\phi}_z}{\partial u}\right)^2\right].  
\end{eqnarray}
Here the instanton configuration $\phi_z$ was defined in
Eq.~(\ref{instanton}).  First term in the right hand side of
Eq.~(\ref{charging1}) describes the charging part of the action on the
instanton configuration and the last two terms describe small fluctuations
around the instanton. Setting $\tilde{\phi}_z=0$ and integrating over the
variable $u$ in Eq.~(\ref{charging1}) we obtain for the charging action on
the instanton configuration,
 \begin{equation}
\label{charginst}
\int^{\beta}_{0} \frac{\dot{\phi}^2_z(\tau)}{4E_C}d\tau
= \frac{\pi^2 T}{E_C}\left(\frac{1+|z|^2}{1-|z|^2}\right).
\end{equation} 
To calculate the second and the third terms in the right hand side of
Eq.~(\ref{charging1}) we expand the massive fluctuations $\tilde{\phi}_z$
using Eq.~(\ref{basis}). Integrating over the variable $u$ in
Eq.~(\ref{charging1}) we obtain the following results
\begin{eqnarray}
\label{second}
&& S_2\equiv \frac{\pi i T}{E_C}\oint \frac{\partial\phi_z}{\partial
  u}\frac{\partial\tilde{\phi}_z}{\partial u} u du = 
\frac{2\pi^2 i T}{E_C(1-|z|^2)}\nonumber \\
&& \times \left(\sum\limits_{n>0}\tilde{\varphi}_n z^{n+1} -
\sum\limits_{n<0} \tilde{\varphi}_n (z^*)^{-n+1}\right), \\  
\label{chargingfluct}
&& S_3 \equiv \frac{\pi i T}{2E_C}\oint
 \left(\frac{\partial\tilde{\phi}_z}{\partial u}\right)^2 
 u du = \frac{\pi^2 T}{E_C}
\nonumber \\ && \times \sum\limits_{n,m>0}\left[\tilde{\varphi}_n
 f_{nm}(z)\tilde{\varphi}_{-m}+ \tilde{\varphi}_m
 f_{mn}(z)\tilde{\varphi}_{-n}\right].  
\end{eqnarray}
Here the function $f_{nm}(z)$ is given by the following expression
\begin{eqnarray}
\label{f}
&& f_{nm}(z) = n^2\delta_{n,m} \nonumber \\ 
&& + (m+n)\left[z^{n-m}\theta(n-m+1) + (z^*) ^{m-n}\theta (m-n)\right]
\nonumber \\ 
&& + z^{n-m}\left[n-m + \frac{1+|z|^2}{1-|z|^2}\right]\theta (n-m+2)
\nonumber \\  
&& + (z^*)^{m-n}\left[m-n + \frac{1+|z|^2}{1-|z|^2}\right]\theta(m-n-1).
\end{eqnarray}
Here again $\theta (0) = 0$.  From Eqs.~(\ref{chargingfluct}) and
(\ref{f}) it follows that the diagonal part of the quadratic form of the
charging action is given by the following expression,
\begin{equation}
\label{diagcharge}
S_3^{diag} = \frac{2\pi^2 T}{E_C}\sum\limits_{n>0}|\tilde \varphi_n|^2
\left[(n+1)^2 + \frac{2|z|^2}{1-|z|^2}\right]. 
\end{equation}
The result of Eq.~(\ref{diagcharge}) was used in Eq.~(\ref{ratio1}).

We now consider the contribution to the renormalized charging energy
$\tilde E_C(T)$ in Eq.~(\ref{result1}) from the term $S_2$ in
Eq.~(\ref{second}). We expand the numerator in the right hand side of
Eq.~(\ref{eq:Zz}) to second order in $S_2$. As a result we obtain the
following correction to the right hand side of Eq.~(\ref{eq:Zzgrabert}), 
\begin{equation}
  \label{correction}
\delta Z^\textrm{G}_1(z)= 
\frac{g_0^2E_C}{2\pi^3 T} \frac{\left\langle S_2^2 \right\rangle_i}{2}
\exp\left[ -\frac{g_0}{2} - 
  \frac{\pi^2 T[1+|z|^2]}{E_C[1-|z|^2]}\right]. 
\end{equation}
Here the angular brackets, $\langle \ldots\rangle_i$, denote averaging
with respect to the Gaussian action, Eq.~(\ref{bare}). From
Eq.~(\ref{second}) we obtain,
\begin{equation}
\label{S2}
\frac{\left\langle S_2^2 \right\rangle_i}{2}=\left(\frac{2\pi^2
    T}{E_C[1-|z|^2]}\right)^2 \sum\limits_{n>0}|z|^{2(n+1)}\left\langle
  |\tilde \varphi_n|^2\right\rangle_i. 
\end{equation}
Using the fact that $\left\langle |\tilde{\varphi}_n|^2 \right\rangle_i =
(g_0 |n|)^{-1}$ and substituting Eqs.~(\ref{correction}), (\ref{S2}) into
Eq.~(\ref{eq:omegaz}) after integration over the $z$ we obtain the
following correction $\delta \tilde E_C(T)$ to the
renormalized charging energy $\tilde E_C(T)$ in Eq.~(\ref{result1})
\begin{equation}
\label{deltaE}
\delta \tilde E_C(T) = \frac{g_0^2E_C}{\pi^2}
\left[\frac{\ln \frac{T_0}{T}}{g_0}\right]\exp\left(-\frac{g_0}{2}\right).
\end{equation}
This correction is proportional to the first power of the large logarithm
$\ln (T_0/T)$. Therefore it is much smaller than the second term in the
right hand side of Eq.~(\ref{result1}) and can be neglected.  It is
straightforward to check that the contribution to the renormalized
charging energy $\tilde E_C(T)$ in Eq.~(\ref{result1}) from the term $S_3$
in Eq.~(\ref{chargingfluct}) is small as $1/g_0^2$ in comparison with the
result of Eq.~(\ref{deltaE}) and therefore can also be neglected.

\section{\label{A} Evaluation of the last three terms in
  Eq.~(\ref{expansion})}

We employ Wick's theorem to rewrite the last three averages in
Eq.~(\ref{expansion}) in the following form,

\begin{eqnarray}
&&\left\langle (\tilde \phi- \tilde \phi_1)^4\right\rangle=3\left\langle
  (\tilde \phi- \tilde \phi_1)^2\right\rangle_i^2, \nonumber \\
&&\left\langle(\tilde \phi-\tilde \phi_1)^3(\tilde \phi_2-\tilde
  \phi_3)^3\right\rangle_i=  6\left\langle(\tilde \phi-\tilde \phi_1)(\tilde
  \phi_2-\tilde \phi_3)\right\rangle_i^3  \nonumber \\  
 &&+ 9\left\langle(\tilde \phi-\tilde \phi_1)(\tilde \phi_2-\tilde
  \phi_3)\right\rangle_i \left\langle(\tilde \phi-\tilde \phi_1)^2
  \right\rangle_i 
  \left\langle(\tilde \phi_2-\tilde 
  \phi_3)^2\right\rangle_i, \nonumber  \\
&&\left\langle ( \phi- \phi_1)^4\right\rangle_0=3\left\langle
  (\phi- \phi_1)^2\right\rangle_0^2. \nonumber 
\end{eqnarray}
Different pairings in this equation can be represented by diagrams in
Fig.~\ref{diagrams}.  

As is verified by the subsequent calculations, only fluctuations of $\phi$
with frequencies below $T/(1-|z|^2)$ contribute to the corresponding
functional integrals.  Therefore we neglect the charging energy term in
the Gaussian action for the massive fluctuations of $\phi$.  It is then
convenient to perform the calculations in the bases
(\ref{eq:matsubarabasis}), (\ref{basis}) which
diagonalize~\cite{Korshunov87} the dissipative action.  For example, for
the correlator $\langle(\tilde \phi-\tilde \phi_1)(\tilde \phi_2-\tilde
\phi_3)\rangle_i$  we obtain the following expression,
\begin{eqnarray}
&&\left\langle(\tilde \phi-\tilde \phi_1)(\tilde \phi_2-\tilde
  \phi_3)\right\rangle_i =  -
2\sum\limits_{n\geq 1}\langle |\tilde \varphi_n|^2\rangle_i  \nonumber
\\  
\label{example}
&& \times {\rm Re}
  \left[\frac{u_1f(u_1)}{u_2f(u_2)}  h_n(u,u_1) h_n(u_2,u_3)
\right].
\end{eqnarray}
Here we introduced the notation 
\begin{equation}
  \label{eq:hdef}
  h_{n}(u,u_1)=1-\left[\frac{u}{u_1}\right]^{n}\frac{f(u)}{f(u_1)}.
\end{equation}
All other correlation functions may be expressed
through $\langle |\tilde \varphi_n|^2\rangle_i$ and $ \langle
|\varphi_n|^2\rangle_0$ in a similar way. Since upon neglecting the charging
action $\langle |\tilde \varphi_n|^2\rangle_i = \langle |\varphi_n|^2\rangle_0
= (g_0 |n|)^{-1}$ we can write the averages in the right hand side of
Eq.~(\ref{expansion}) in the following form:
\begin{widetext}
\begin{subequations}
\label{eq:suint}
\begin{eqnarray}
\label{eq:suinta}
\left\langle S^{i}_{\tilde \phi^4}\right\rangle_i &=&\frac{2}{g_0} 
\sum\limits_{n,n_1\geq 1}^{\infty}\frac{1}{nn_1} 
\oint\frac{dudu_1}{(2\pi i)^2(u-u_1)^2} {\rm Re}
\left[\frac{f(u)}{f(u_1)}\right] 
 {\rm Re}\left[h_n(u,u_1)\right]{\rm
  Re}\left[h_{n_1}(u,u_1)\right], \\ 
\label{eq:suintb}
\left\langle S^{0}_{\phi^4}\right\rangle_0 &=&\frac{2}{g_0}
\sum\limits_{n,n_1\geq 2}^{\infty}\frac{1}{nn_1}
\oint\frac{dudu_1}{(2\pi
  i)^2(u-u_1)^2}   
{\rm Re}\left[1-\left[\frac{u}{u_1}\right]^{n}\right]{\rm
  Re}\left[1-\left[\frac{u}{u_1}\right]^{n_1}\right], \\ 
\frac{\left\langle (S^i_{\tilde\phi^3})^2\right\rangle_i}{2} &=&
 - \frac{2}{3g_0} \sum\limits_{n,n_1,n_2\geq  
  1}^{\infty}\frac{1}{n n_1 n_2}\oint\frac{du d u_1 d u_2 d u_3
  }{(2\pi)^4(u-u_1)^2(u_2-u_3)^2}{\rm  Im}\left[\frac{f(u)}{f(u_1)}\right]
\nonumber \\ 
\label{eq:suintc}
&& \times \Big( 6 {\rm Re}[h_{n_1}(u,u_1)]{\rm
  Re}[h_{n_2}(u_2,u_3)]{\rm Re}[y_n]  
+ {\rm Re}[y_n]{\rm Re}[y_{n_1}]{\rm Re}[y_{n_2}]\Big), 
\end{eqnarray}
\end{subequations}
\end{widetext}
where $y_{n_i}$ denotes the following function,
\begin{equation}
  \label{eq:ydef}
    y_{n_i}=\left[\frac{u_1}{u_2}\right]^{n_i}
    \frac{f(u_1)}{f(u_2)}h_{n_i}(u,u_1)h_{n_i}(u_2,u_3).  
\end{equation}
The two terms in the last line of Eq.~(\ref{eq:suintc}) correspond to the
second and third diagrams in Fig.~\ref{diagrams} respectively.

Next we perform the integration over $u_i$ in Eq.~(\ref{eq:suint}).  The
first term in the last line of Eq.~(\ref{eq:suintc}) vanishes. 
The remaining integrals lead to the results:
\begin{subequations}
\label{eq:spertresult}
\begin{equation}
\label{eq:spertresulta}
\left\langle
  S^{i}_{\tilde\phi^4}\right\rangle_i=-\frac{1}{2g_0}\sum\limits_{n,n_1\geq 
  1}^{\infty}\frac{2{\rm
    min(n,n_1)}-|z|^{2|n-n_1|}}{n n_1}, 
\end{equation}
\begin{eqnarray}
\label{eq:spertresultb}
&& \frac{\left\langle (S^i_{\tilde \phi^3})^2\right
 \rangle_i}{2}=-\frac{[1-|z|^2]^2}{6g_0} 
 \sum\limits_{n,n_1,n_2\geq 1}^{\infty}\frac{1}{n n_1 n_2} \\ 
&&  \times [2 (n_2-n-n_1)^2|z|^{2(n_2-n_1-n-1)}\theta(n_2-n-n_1) 
\nonumber \\
&& +  (n-n_2-n_1)^2|z|^{2(n-n_1-n_2-1)}\theta(n-n_2-n_1)], \nonumber \\
\label{eq:spertresultc}
&&\left\langle S^{0}_{\phi^4}\right\rangle_0 = -\frac{1}{g_0}
\sum\limits_{n,n_1\geq 1}^{\infty}\frac{{\rm min(n, n_1)}}{n n_1}.
\end{eqnarray}
\end{subequations}
Evaluating the sums in Eq.~(\ref{eq:spertresult}) with logarithmic accuracy
and using Eq.~(\ref{expansion}) we obtain Eq.~(\ref{eq:Zzpert}).

\end{document}